# Title: Stabilizing sample-wide Kekulé orders in graphene/transition metal dichalcogenide heterostructures

**Authors:** Mo-Han Zhang[§], Ya-Ning Ren[§], Qi Zheng[§], Xiao-Feng Zhou, Lin He[†]

**Affiliations:**
Center for Advanced Quantum Studies, Department of Physics, Beijing Normal University, Beijing, 100875, People's Republic of China

[§]These authors contributed equally to this work.
[†]Correspondence and requests for materials should be addressed to Lin He (e-mail: helin@bnu.edu.cn).

**Kekulé phases are Peierls-like lattice distortions in graphene that are predicted to host novel electronic states beyond graphene (*1-8*). Although the Kekulé phases are realized in graphene through introducing electron-electron interactions at high magnetic fields (*9-11*) or adatom superlattices (*12-15*), it is still an extremely challenge to obtain large-area graphene Kekulé phases in experiment. Here we demonstrate that sample-wide Kekulé distortions in graphene can be stabilized by using transition metal dichalcogenides (TMDs) as substrates and the induced Kekulé orders are quite robust in the whole graphene/TMDs heterostructures with different twist angles. The commensurate structures of the heterostructures provide periodic scattering centers that break the translational symmetry of graphene and couple electrons of the two valleys in graphene, which tips the graphene toward global Kekulé density wave phases. Unexpectedly, three distinct Kekulé bond textures stabilized at various energies are directly imaged in every graphene/TMDs heterostructure. Our results reveal an unexpected sensitivity of electronic properties in graphene to the supporting substrates and provide an**

**attractive route toward designing novel phases in graphene/TMDs heterostructures.**

Graphene is well-known for its rigidity of the honeycomb lattice that exhibits sixfold bond symmetry (*16,17*). However, it was predicted in theory that a bond density wave instability, i.e., the Kekulé distortion that breaks the bond symmetry and triples the unit cell, can be triggered in graphene through several different mechanisms (*1-8*). Recently, two types of the Kekulé phases are explicitly demonstrated through direct experimental observations. One is Kekulé-O order with "O"-shaped bond modulation that can be induced by electron-electron interactions at high magnetic fields (*9-11*) or by adatom superlattices (*12-14*). The other is Kekulé-Y order, where modulated bonds form a "Y"-shape pattern, that is realized in graphene grown epitaxially onto Cu(111), in which one of six carbon atoms in each superlattice unit cell is on top of a copper monovacancy (*15*). At an experimental level, it is highly challenging to stabilize the Kekulé orders in graphene over large areas in zero magnetic field because the requirement of precise control of the adatoms or monovacancies' positions (*12-15*). Here, we show that superposing graphene on two-dimensional (2D) transition metal dichalcogenides (TMDs), $WSe_2$ and $WS_2$, creates superlattices that generate the Kekulé phases over the whole graphene/TMDs heterostructures. The commensurate structures of the heterostructures provide periodic scattering centers to couple electrons of the two valleys and generate global Kekulé density wave phases in graphene. In each graphene/TMDs heterostructure, three different Kekulé bond textures, which are stabilized at different energies, are directly imaged.

Figure 1a shows real-space schematics of four bond textures, *i.e.*, Kekulé-Y-1, Kekulé-Y-2, Kekulé-O, and Kekulé-M, in graphene (*1-15,18*). In all the Kekulé phases, the unit cells are tripled in size. However, the bond textures of them are remarkably different, which enables us to distinct them by directly imaging the C-C bond density wave in real space (*1-8*). For the Kekulé-Y phase, there are two (four) bonds among the six bonds in each honeycomb exhibiting high density in the Kekulé-Y-1 (Kekulé-Y-2) bond texture. For the Kekulé-O phase, the "O"-shaped bond modulation is easy to identify in the real-space image. For the Kekulé-M phase, all bond strengths are equal, however, there are three different on-site energies. To date, only the Kekulé-Y-1 and the Kekulé-O bond textures are directly imaged in experiment (*9-15*). Instead of using electron-electron interactions at high magnetic fields or precisely controlling positions of adatoms, here we propose to create global Kekulé phases in graphene through moiré superlattices from stacked and twisted structures of graphene on 2D TMDs, as shown in Fig. 1b. The atomic sites of the TMDs provide superlattice of scattering centers to generate wave vectors that satisfy the Kekulé nesting condition $\delta q = K - K'$ (the wave vector connects two Brillouin zone corners, *i.e.*, the intervalley scattering). Once the Friedel oscillations generated by these scattering centers are in phase, the charge-density modulations are enhanced and, then, sample-wide distinct Kekulé bond density phases are stabilized in graphene.

To explore possible Kekulé phases in the graphene/TMDs heterostructures, different samples with controlled twist angles are obtained by using transfer technology of graphene monolayer onto mechanical-exfoliated $WSe_2$ and $WS_2$ sheets (*19-21*) (See Fig. 2a and methods for details of the sample preparation). The obtained graphene/TMDs heterostructures are systematically characterized by carrying out scanning tunneling microscope (STM) and scanning tunneling spectroscopy (STS) measurements (see Supplemental Fig. 1) and, unexpectedly, the Kekulé phases are quite robust for most of the studied heterostructures. Figure 2b shows a representative STM image and a STS map of a graphene/WSe$_2$ heterostructure with a rotation angle 49.3° between graphene and the WSe$_2$. The first characteristic of the Kekulé distortion is clearly shown by the presence of a commensurate $\sqrt{3} \times \sqrt{3}$ $R$ 30° charge-density wave (CDW) in both the STM image and the STS map, which directly reflects the local density of states at the measured energy. The obtained $\sqrt{3} \times \sqrt{3}$ $R$ 30° CDW order is long range over the whole sample when the stacked structure of the heterostructure is uniform. In our experiment, we obtain similar CDW order in different areas of the 49.3° graphene/WSe$_2$ heterostructure. The atomic-resolved STM measurements and the sample-wide uniform CDW order also help us to completely exclude atomic defects or adatoms as the origin of the observed Kekulé distortion. The emergence of long-range Kekulé distortion in the graphene honeycomb lattice is further confirmed by Fourier transforms (FT) of large-area topographic image. In the FT image, beside the reciprocal lattice of graphene, the reciprocal lattice of WSe$_2$, and the reciprocal lattice of moiré structure in the graphene/WSe$_2$ heterostructure, we can observe sharp peaks at $Q_{Kek}$, arising from the $\sqrt{3} \times \sqrt{3}$ $R$ 30° CDW (Figs. 2c and 2d), as observed in the graphene Kekulé phase in previous study (*15*).

In our experiment, direct evidence of the Kekulé bond order in graphene is obtained by performing high-resolution STS measurements. Figure 2e(right middle panel) shows a representative STS map of the 49.3° graphene/WSe$_2$ heterostructure, which clearly demonstrates strongly modified topography, with a periodic arrangement of highly distorted hexagons compatible with the Kekulé-Y-1 type pattern. Obviously, a third of the bonds in graphene are quite different due to the formation of the Kekulé density wave phase. It is interesting to note that the Kekulé-Y-1 type pattern is frequently observed in the graphene/WSe$_2$ heterostructures with different rotation angles, as shown in Fig. 2e. In the presence of magnetic fields, Landau quantization of massless Dirac fermions is detected in the graphene/WSe$_2$ heterostructures (see supplemental Fig. S2), indicating that there is still zero gap in the graphene Kekulé phase, which is the first experimental demonstration of such a theoretical prediction (*1-8*). To demonstrate the universality of the mechanism in introducing the Kekulé phases in graphene by using 2D TMDs as the supporting substrate, we carry out similar measurements in graphene/WS$_2$ heterostructures. The same Kekulé-Y-1 type pattern is also frequently observed in graphene/WS$_2$ heterostructures with different rotation angles, as shown in Fig. 2e (see supplemental Fig. S3 for more experimental evidences), which explicitly demonstrates the universality of the proposed mechanism.

When a 2D crystal with a lattice constant $a$ is twisted $\theta$ relative to another with a lattice mismatch of $\delta$, the moiré length is given by $\lambda = \frac{(1+\delta)a}{\sqrt{2(1+\delta)(1-\cos\theta)+\delta^2}}$ (*22-24*). With considering the lattice constants of graphene and WSe$_2$ (WS$_2$), the maximum moiré length generated between them is 0.984 nm (1.123 nm). Usually, it is believed that such a small moiré pattern generated between graphene and 2D insulating substrates will not affect the structure and electronic properties of graphene. Our result is contrary to such a belief and demonstrates that the 2D TMDs introduce robust Kekulé phases in graphene. To introduce large-area Kekulé phases in graphene, the Friedel oscillations generated by the scattering centers in the supporting substrate must be in phase (*12-15*), which requires that the scattering centers occupy one of the red-green-blue (RGB) sites of graphene, as schematically shown in Fig. 3a. In the RGB mosaic representation of the graphene lattice, the size of the unit cell is tripled and the Kekulé phases of graphene can be represented by making a three-colour tiling of the pristine graphene lattice (*12-15*). In the graphene/TMDs heterostructures, a stacked commensurate structure naturally provides atomic structures as the periodic scattering centers. When all the periodic scattering centers in the commensurate structure occupy the same RGB sites, the interference between scattering-induced oscillations of different centers is constructive (Fig. 3a, right panel), inducing the Kekulé distortion in graphene globally. Here we should point out that the length scale of the commensurate structure is not necessary the same as that of the unit cell of the Kekulé phase, actually, it could be much larger than that of the unit cell. Very recently, it has demonstrated explicitly that the Kekulé phase can be induced in graphene through 0.3% surface coverage lithium adatoms, corresponding to an average Li-Li distance of about 2 nm (*16*). Figure 3b shows the calculated length scale of the commensurate structure in the graphene/$WSe_2$ heterostructures as a function of the twist angles. In the calculation, small lattice strain is assumed in graphene to create commensurate supercells (*24-27*) and a restriction on inducing the Kekulé distortion in graphene is imposed on the length scale of the commensurate structure (see Fig. S4 and supplemental materials for details of calculation). In graphene, the de Broglie wavelength of massless Dirac fermions is quite large around the Dirac point (for example, it is easy to reach several hundreds nanometers around the Dirac point and it is still as large as 10 nm at 400 meV away from the Dirac point) (*28*). Therefore, each scattering center can nucleate a local patch of Kekulé order of a quite large radius, as reported around each adatom previously (*14,28*). The interactions between the scattering centers may prefer an arrangement that allows their Friedel oscillations to be in phase and, consequently, prefer the formation of commensurate structures in the graphene/TMDs heterostructures. Here, we assume a boundary of less than 10 nm on the length scale of the commensurate structure to stabilize the Kekulé phase. Obviously, the Kekulé phase is expected to be observed in a wide range of twist angles in the graphene/TMDs heterostructures, as shown in Fig. 3b. Therefore, we can frequently observe the Kekulé phase in the heterostructures in our experiment.

Besides the Kekulé-Y-1 order, as shown in Fig. 2, the other two bond textures of the Kekulé phases, *i.e.*, the Kekulé-Y-2 and Kekulé-M, are also observed in all the studied graphene/TMDs heterostructures. Figure 4 summarizes representative results obtained in the 20.6° graphene/WSe$_2$ heterostructure and in the 15.6° graphene/WS$_2$ heterostructure, respectively (see supplemental Figs. S5 and S6 for more experimental results). In previous studies, usually, only one type of the Kekulé distortion can be observed in a studied system (*9-15*). Emergence of three bond textures of the Kekulé phases in each single graphene/TMDs heterostructure is an unexpected result in our experiment. According to our experiment, the three bond textures of the Kekulé phases emerge at different energies of the graphene. The three Kekulé bond textures seem appear randomly at different energies in different heterostructures and there is no obvious energy dependence of the Kekulé bond textures (Fig. 4 and Fig. S5). Besides emerging at different energies, the vertexes of the unit cells for the three Kekulé bond textures could also be different, as shown in Fig. 4 and Fig. S5. In the 20.6° graphene/WSe$_2$ heterostructure, the vertexes of the unit cells for the Kekulé-Y-1 and Kekulé-M are the same, however, there is a shift of C-C distance from that of the Kekulé-Y-2 bond texture, as shown in Fig. 4e. In the 15.6° graphene/WS$_2$ heterostructure, the vertexes of the unit cells for the Kekulé-Y-1 and Kekulé-Y-1 are the same, however, there is a shift of 0.246 nm from that of the Kekulé-M bond texture, as shown in Fig. 4f. These results indicate that the three Kekulé bond textures may be generated by different scattering centers in the commensurate structure. In the commensurate structure, every atomic sites of the 2D TMDs can act as the scattering centers for the Kekulé phases since that a translational symmetry is retained. Then, different scattering centers stabilize the graphene in different Kekulé bond textures at various energies.

In summary, we demonstrate the ability to introduce sample-wide Kekulé phases in graphene by using TMDs as substrates. The commensurate structures of the graphene/TMDs heterostructures provide periodic scattering centers that tip the graphene toward global Kekulé density wave phases. Three Kekulé bond textures stabilized at various energies are directly imaged in every single graphene/TMDs heterostructure. Our results provide an attractive route toward designing novel phases in graphene by using 2D TMDs as supporting substrates.

**Acknowledgments:**

This work was supported by the National Key R and D Program of China (Grant Nos. 2021YFA1401900, 2021YFA1400100) and National Natural Science Foundation of China (Grant Nos.12141401, 11974050).

**Author contributions**

M.H.Z., Y.N.R., Q.Z. and X.F.Z. performed the sample synthesis, characterization and STM/STS

measurements. M.H.Z. and L.H. analyzed the data. L.H. conceived and provided advice on the experiment, analysis and the theoretical calculations. M.H.Z. and L.H. wrote the paper with the input from others. All authors participated in the data discussion.

**Data availability statement**

All data supporting the findings of this study are available from the corresponding author upon request.

**Methods**

**Sample preparation.** The studied devices were fabricated using a wet-/dry-transfer technique. First of all, the TMDs crystal was separated into thick-layer $WSe_2$ (or $WS_2$) sheets by scotch tape. We stacked thick-layer $WSe_2$ (or $WS_2$) sheets to 285 nm thick $SiO_2$/Si with polydimethylsiloxane (PDMS). The Si in the bottom is highly N-doped. Then, we grow large area graphene monolayer films on a $20 \times 20$ mm$^2$ polycrystalline copper (Cu) foil (Alfa Aesar, 99.8% purity, 25 μm thick) via a low-pressure chemical vapor deposition (LPCVD) method. The Cu foil was heated from room temperature to 1035 ºC in 30 min and annealed at 1035 ºC for ten hours with Ar flow of 50 sccm and $H_2$ flow of 50 sccm. Then $CH_4$ flow of 5 sccm and the other gases to be the same as before was introduced for 20 min to grow high-quality large area graphene monolayer. The furnace was cooled down naturally to room temperature. We used wet-transfer technique with polymethyl methacrylate (PMMA) to transfer graphene monolayer onto the substrate (TMD/$SiO_2$/Si). Finally, we cleaned the upper PMMA with acetone.

**Measurements.** STM/STS measurements were performed in low-temperature (4.2 K) and ultrahigh-vacuum (~$10^{-10}$ Torr) scanning probe microscopes [USM-1300 (4.2 K)] from UNISOKU. The tips were obtained by chemical etching from a W (95%) alloy wire. Before the experiment, we calibrate the STM with Highly Oriented Pyrolytic Graphite. The differential conductance (d$I$/d$V$) measurements were taken by a standard lock-in technique with an ac bias modulation of 5 mV and 793 Hz signal added to the tunneling bias.

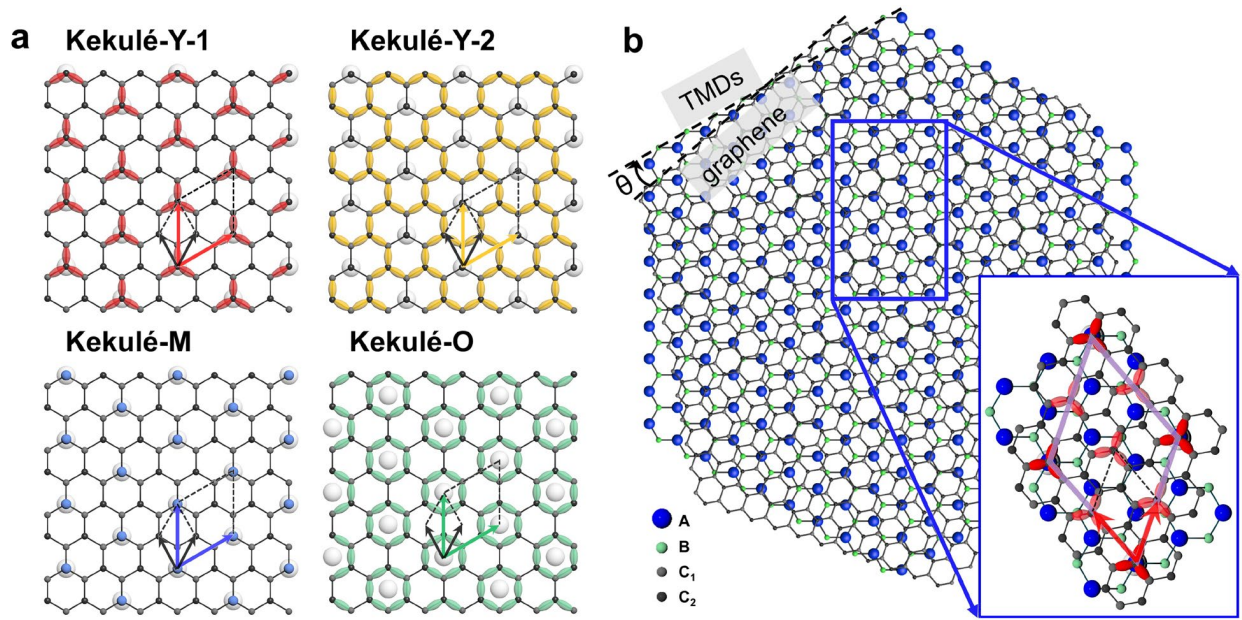

**Fig. 1 | The schematic of Kekulé orders and a graphene/TMDs heterostructure. a**, A schematic for four Kekulé bond textures, showing different bond strengths or atomic on-site energies, in graphene. The unit cells of pristine graphene and Kekulé-ordered graphene are marked by black and colored parallelograms, respectively. **b**, A schematic drawing of graphene/TMDs heterostructure with a rotation angle $\theta$. The inset shows the zoom-in atomic structure in blue frame. The purple diamond represents the unit cell of the commensurate structure in the heterostructure. That atomic sites at vertexes of the purple diamond can provide periodic scattering centers to induce long-range Kekulé order in graphene.

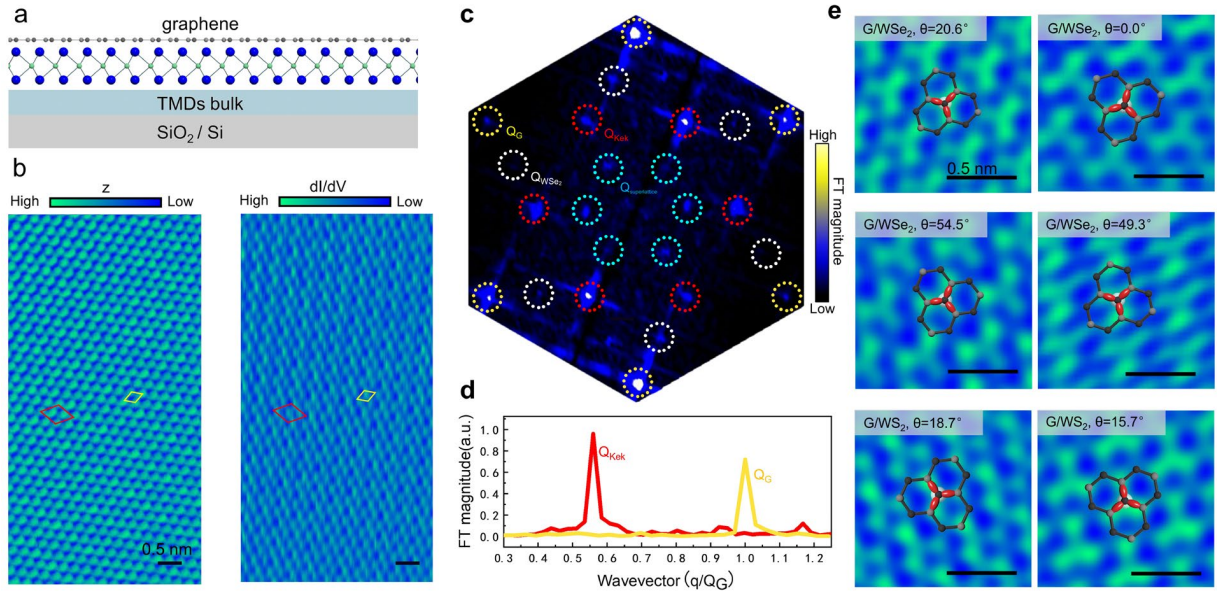

**Fig. 2 | Large-scale Kekulé order in graphene/TMDs heterostructures. a.** Device structure. Monolayer graphene is placed on 2D TMDs, then the graphene/TMDs heterostructure is placed on SiO$_2$/Si substrate. **b.** STM image and STS map (measured at 40 meV) in a graphene/WSe$_2$ heterostructure with twisted angle 49.3° between graphene and the WSe$_2$. The unit cells of Kekulé order and graphene lattice are marked by red and yellow diamonds, respectively. **c.** The FT of the STS map in panel **b**. The yellow, white, green circles are the graphene, WSe$_2$, graphene/WSe$_2$ moiré Bragg peak locations, respectively. The six red circles are at the wavelength of the Kekulé order. **d.** Line cuts through the graphene and Kekulé order peaks in the FT in **c**. The Kekulé order peak is as sharp as the graphene peak. **e.** Atomic resolved STS maps showing the Kekulé-Y-1 bond texture in six different graphene/TMDs heterostructures. The measured energies are 167 meV (G/WSe$_2$, $\theta = 20.6°$), 225 meV (G/WSe$_2$, $\theta = 0.0°$), 38 meV (G/WSe$_2$, $\theta = 54.5°$), 223 meV (G/WSe$_2$, $\theta = 49.3°$), 80 meV (G/WS$_2$, $\theta = 18.7°$), and -8 meV (G/WS$_2$, $\theta = 15.7°$).

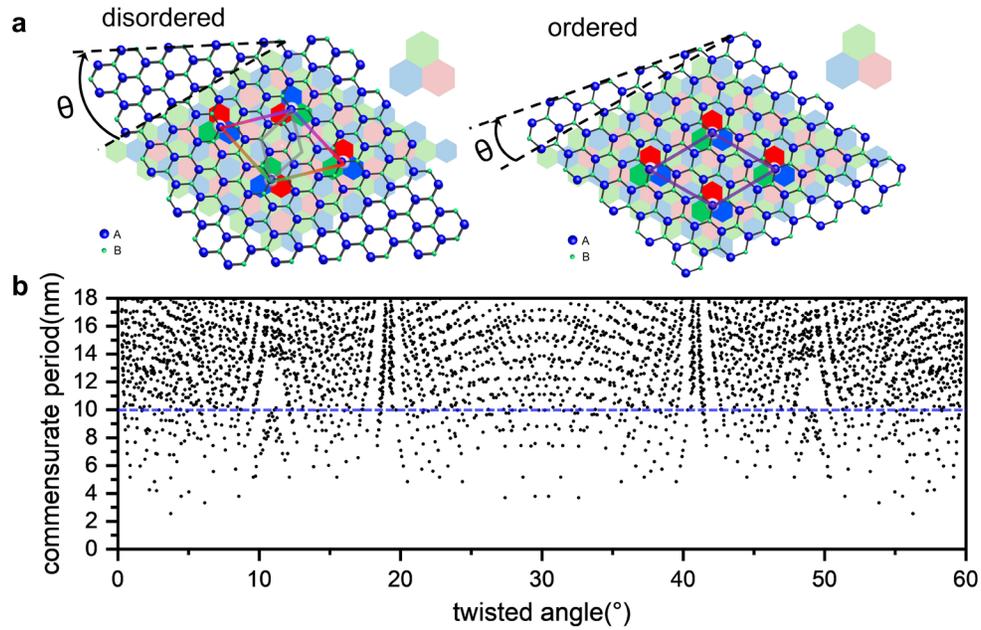

**Fig. 3 | Kekulé ordered states in the commensurate structures of graphene/TMDs heterostructures. a.** Illustrations of the disordered and ordered Kekulé states of the commensurate structures. The Kekulé phase of graphene is represented by making a red-blue-green (RBG) mosaic tiling of the pristine graphene lattice. Atomic structure of the TMD is overlaid onto the Kekulé phase of graphene. The unit cell of the commensurate structure is marked by chromatic parallelogram. Large-scale Kekulé ordered state is formed when the vertices of the commensurate structure occupy the same RGB sites. When they occupy different RGB sites, the induced oscillations by different scattering centers are not in phase and we cannot observe the Kekulé distortion globally. **b.** Calculated commensurate periods as a function of twist angle in the graphene/WSe$_2$ heterostructures. In the calculation, a restriction on inducing the large-scale Kekulé distortion in graphene is imposed on the length scale of the commensurate structure. A boundary of about 10 nm on the length scale of the commensurate structure is assumed to stabilize the Kekulé phase in graphene.

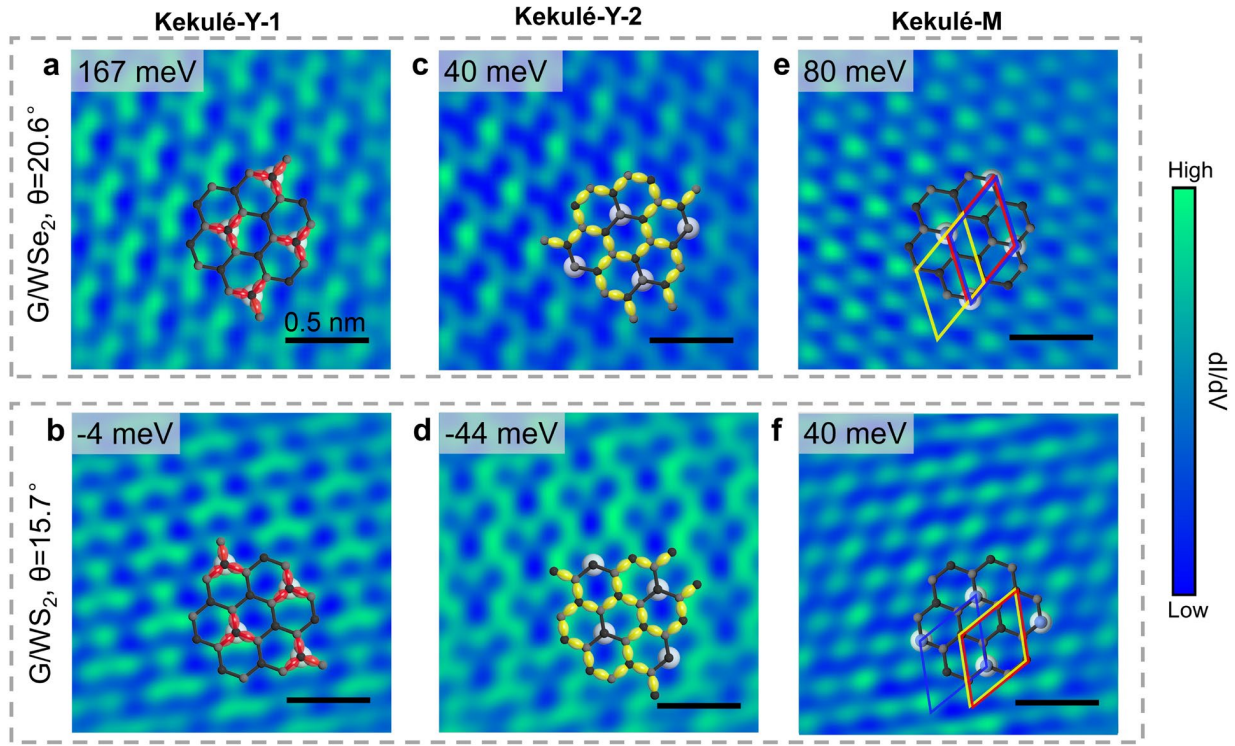

**Fig. 4 | Different Kekulé bond textures in graphene/TMDs heterostructures. a,c,e.** Kekulé-Y-1, Kekulé-Y-2 and Kekulé-M orders observed in a 20.6° graphene/WSe$_2$ heterostructure. **b,d,f.** Kekulé-Y-1, Kekulé-Y-2 and Kekulé-M orders observed in a 15.6° graphene/WS$_2$ heterostructure. The unit cells of the Kekulé bond textures are marked by red (Kekulé-Y-1), yellow (Kekulé-Y-2), and blue (Kekulé-M) parallelograms in panels e and f.